\documentclass[prl,english,twocolumn,amsfonts,amssymb,floats,showpacs,aps]{revtex4-1}

\usepackage[final,dvips]{epsfig}
\usepackage{amsmath,amsfonts}
\usepackage{bbm}
\usepackage{float}
\usepackage[caption = false]{subfig}
\usepackage{graphicx}
\usepackage{color}
\usepackage{hyperref}
\usepackage{natbib}
\usepackage{multibib}

\newcommand{\jh}{J} % NN Heisenberg coupling 
\newcommand{\jk}{K} % Kitaev interaction
\newcommand{\gm}{\Gamma} % Bond anisotropy
\newcommand{\hz}{\vec{h}} % External field

\newcommand{\bc}{b^{\dagger}} % Boson creation
\newcommand{\ba}{b} % Boson annihilation

\newcommand{\bt}{\beta} % Bogoliubov quasiparticle

\newcommand{\ii}{\iota} % Math symbol for \sqrt(-1)
\newcommand{\kk}{\vec{k}} % Momentum vector

\begin{document}

\title[]
{
Topological excitations in the ferromagnetic Kitaev-Heisenberg model
}
\author{Darshan G. Joshi}
\email[]{d.joshi@fkf.mpg.de}
\affiliation{Max-Planck-Institute for Solid State Research, D-70569 Stuttgart, Germany}

\date{\today}

\begin{abstract}
With the advancement in synthesizing and analyzing Kitaev materials, the Kitaev-Heisenberg model on the honeycomb lattice has attracted a lot of attention in the last few years. Several variations, which include additional anisotropic interactions as well as response to external magnetic field, have been investigated and many exotic ordered phases have been discussed. On the other hand, quantum spin systems are proving to be a fertile ground to realize and study bosonic analogues of fermionic topological states of matter. Using the spin-wave theory we show that the ferromagnetic phase of the extended Kitaev-Heisenberg model hosts topological excitations. Along the zig-zag edge of the honeycomb lattice we find chiral edge states, which are protected by a non-zero Chern number topological invariant. We discuss two different scenarios for the direction of the spin polarization namely $[001]$ and $[111]$, which are motivated by possible directions of applied field. Dynamic structure factor, accessible in scattering experiments, is shown to exhibit signatures of these topological edge excitations. Furthermore, we show that in case of spin polarization in $[001]$ direction, a topological phase transition occurs once the Kitaev couplings are made anisotropic. %Experimental signatures via dynamical spin correlations are also discussed. 
\end{abstract}

%\pacs{ }

\maketitle

%%%%%%%%%%%%%%%%%%%%%%%%%%%%%%%%%%%%%%%%%%%%%%%%%%%%%%%%%%%%%%%%%%%%%%%%%
%%%%%%%%%%%%%%%%%%%%%%%%%%%%%%%%%%%%%%%%%%%%%%%%%%%%%%%%%%%%%%%%%%%%%%%%%

\emph{Introduction.--} 
Topological properties of electronic energy bands in solid-state systems has ushered in a new paradigm of Physics \cite{hasan_kane, qi_zhang}.
%\todo{cite several topo papers}. 
Several exotic states of matter have been able to be identified with the help of this new tool. Over the last decade these have been systematically classified using symmetry principles \cite{classif1, classif2}. 
Fermionic topological states of matter have a distinct ground state as compared to their trivial counterpart. There are localized edge states in these systems and much efforts have been invested in detecting these signatures. 

However, topological properties need not be restricted to ground state. Energy bands in an excitation spectrum can also possess non-trivial topology and thus lead to topologically non-trivial excitations. This situation is particularly interesting because a ground state property can not distinguish a system with or without these topological excitations. It is therefore an interesting avenue also for experiments to develop new ways to detect these exotic topological systems. 

There have been several examples of bosonic systems hosting topological excitations 
\cite{shindou_murakami_PRB_13,zhang_li_magnonics_PRB_13,peano_marquardt_PRX_15,joannopoulos_review_photonics,cold_atm1,cold_atm2}.
Of particular interest are quantum spin systems because most of them host bosonic quasiparticle excitations. There have been some proposals already wherein the magnetically ordered phases \cite{ss_hall,ss_hall_exp,kps_ss,weyl_magnon,sk_kim,owerre1,owerre2} as well as quantum paramagnetic phase \cite{tqp} host topological edge excitations. 
 
Frustrated magnets \cite{balents,mv_review} are a rich playground for exotic states of matter. %\todo{cite?}. 
One of the examples of intense recent research is the Kitaev model \cite{kitaev}, which is exactly solvable and hosts a 
$\mathbbm{Z}_{2}$ quantum spin liquid. While there is no established example of a material having only Kitaev model, there are some candidate materials, like Na$_2$IrO$_3$ and $\alpha-$RuCl$_3$,whose ordered phases indicate presence of Kitaev interaction along with the Heisenberg exchange \cite{aRu,Na1,Na2}.
This has led to the study of the Kitaev-Heisenberg model \cite{JK_1,kh_sh}, which is a subject of intense research in the last few years (see for instance, Refs. \cite{JK_2,kh_ex1,kh_ex2,kh_field,kh_field2,kh_dynamics,kh_obd,kitaev_topo} ) .

The Kitaev-Heisenberg model with additional symmetry allowed spin-anisotropic terms has also been proposed \cite{kh_ex1,kh_ex2} to understand the experimental data of several candidate materials hosting a Kitaev interaction. While it is difficult to choose a particular model with given interactions to explain the experiments, it has been suggested that application of an external field provides a good handle to restrict the choices \cite{kh_field,kh_field2}. 

Given the rich Physics arising in the Kitaev-Heisenberg model, it is an important question to ask whether the possible ordered  phases could host topological excitations. Especially with the vicinity of the Kitaev spin liquid, this question becomes highly relevant. In this work we answer this question positively by showing that the ferromagnetic phase of the Kitaev-Heisenberg model on a honeycomb lattice hosts bosonic topological excitations in presence of a spin anisotropic interaction. In particular, we show that it has topologically protected chiral edge states and a topological phase transition can be tuned by applying an external magnetic field or making the Kitaev couplings anisotropic.

\begin{figure}
\centering 
\includegraphics[width=0.3\textwidth]{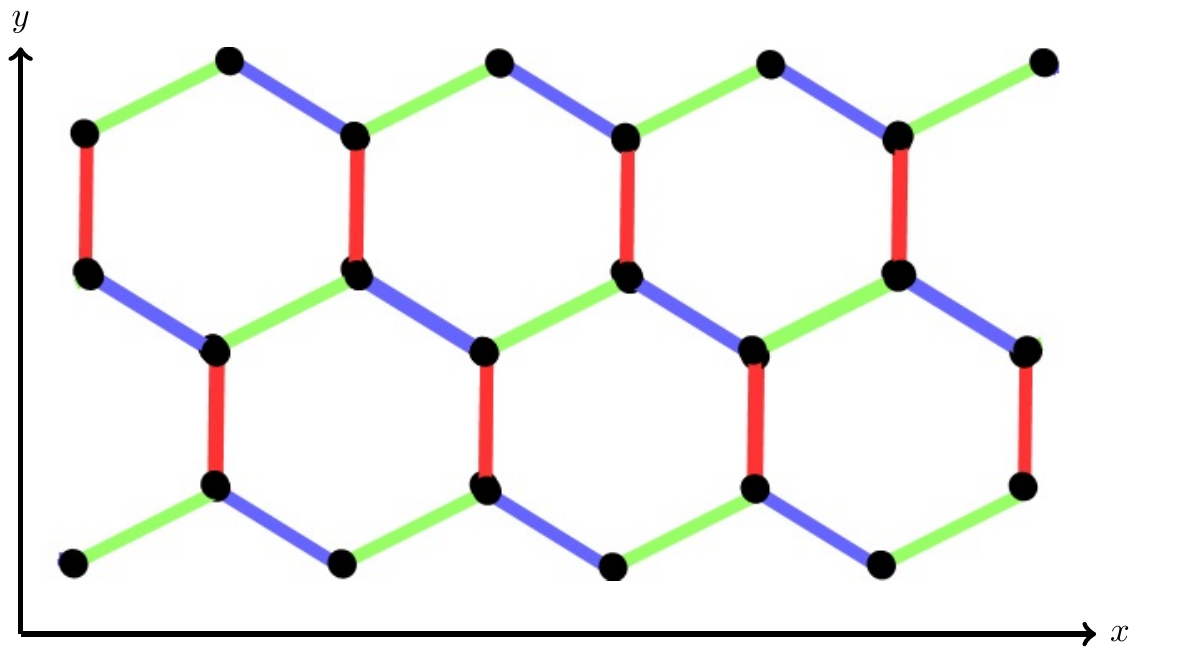}
\caption{Extended Kitaev-Heisenberg model on the honeycomb lattice. The black dots are spins with $S=1/2$ at the honeycomb lattice sites. Each nearest neighbor spin interacts via the Heisenberg exchange. The colors on the bonds represent the bond-dependent Kitaev interaction; $\lbrace$ blue, green, red $\rbrace$ $\rightarrow$ $\lbrace \jk^{x}, \jk^{y}, \jk^{z} \rbrace$. In addition, we also consider the spin-anisotropic exchange as discussed in the text. The upper and the lower edge (running along the $x-$direction) is the so called zig-zag edge.  }
\label{fig:model}
\end{figure}

\emph{Model.--} 
We consider the extended Kitaev-Heisenberg model, i.e., symmetry allowed spin-anisotropic interaction in addition to the Kitaev and Heisenberg exchange terms \cite{JK_1,JK_2,kh_ex1,kh_ex2} on a honeycomb lattice (see Fig. \ref{fig:model}). An external magnetic field is also considered. The corresponding Hamiltonian is written as follows:
\begin{align}
\label{eq:khg}
\mathcal{H} &= \jh \sum_{\langle ij \rangle} \vec{S}_{i} \cdot \vec{S}_{j} 
+ 2 \sum_{\langle ij \rangle_{\gamma}} \jk^{\gamma} S_{i}^{\gamma} S_{j}^{\gamma}  \nonumber \\
&+ \sum_{\langle ij \rangle_{\gamma}} \gm^{\gamma} \left[ S_{i}^{\alpha} S_{j}^{\beta} 
+ S_{i}^{\beta} S_{j}^{\alpha} \right] %\nonumber \\
- \hz \cdot \sum_{i} \vec{S}_{i} \,,
\end{align}
 where the first term is the nearest-neighbor Heisenberg exchange, the second term is the bond-dependent Kitaev interaction, the third term is the bond-dependent spin-anisotropic interaction, and the last term is the coupling to external magnetic field. If one parametrizes $\jh=\cos\phi$ and $\jk=\sin\phi$ ($\jk^{\gamma}=\jk$) then it is known that for $\gm=\hz=0$ a ferromagnetic phase exists in the region $0.85\pi < \phi < 3\pi/2$ \cite{kh_field}. This phase survives even in the presence of small $\gm$ interaction. We shall be primarily interested in this phase.

%%%%%%%%%%%%%%%%%%%%%%%%%%%%%%%%%%%%%
%%%%%%%%%%%%%%%%%%%%%%%%%%%%%%%%%%%%%

\emph{Spin-wave theory.--}
The elementary excitations of the ferromagnetic phase are spin waves,  with the corresponding quasiparticle called as magnons. These are best studied within the spin-wave theory by representing the spin operators in terms of auxiliary bosons via the Holstein-Primakoff transformation \cite{hp}: $S^{z}=S-\bc\ba, S^{+}=\sqrt{2S-\bc\ba}\ba, S^{-}=\bc\sqrt{2S-\bc\ba}$; $\bc$ being the boson creation operator. Here $1/S$, where $S$ is the spin magnitude, is used as a systematic expansion parameter and the resulting spin-wave Hamiltonian contains terms arranged in powers of $1/S$ \cite{footnote_swt}. We shall restrict ourselves to the linear spin-wave theory wherein we consider only bilinear terms in the bosonic Hamiltonian. Such an approximation is controlled and is known to work remarkably well in predicting the correct Physics \cite{footnote_lswt}. %{\clr even in highly frustrated systems (see for instance, Refs. \cite{sq,tri})}. % \todo{cite exmaples}. 
Upon inserting the above Holstein-Primakoff transformation into the Hamiltonian in Eq. \ref{eq:khg} and after subsequent Fourier transformation, we obtain the following linear spin-wave theory Hamiltonian:
%The linear spin-wave theory Hamiltonian can be written as follows:
\begin{equation}
\label{eq:h2k}
\mathcal{H}_{2\kk}=\frac{S}{2} \sum_{\kk} \Psi^{\dagger}_{\kk} \mathcal{M}_{k} \Psi_{\kk} \,,
\end{equation}
where $\Psi_{\kk}=\left( \ba_{A,\kk}, \ba_{B, \kk}, \bc_{A,-\kk}, \bc_{B,-\kk} \right)^{T}$ with $\bc_{A,B}$ being the magnon creation operator on sublattice A (B). The matrix $\mathcal{M}_{\kk}$ is of the form,
\begin{equation}
\label{eq:mk_gen}
\mathcal{M}_{\kk} =
\left(
\begin{matrix}
A_{\kk} & B_{\kk} \\
B^{\dagger}_{\kk} & A^{T}_{-\kk}
\end{matrix}
\right) \,,
\end{equation}
where $A_{\kk}$ and $B_{\kk}$ are momentum-dependent functions of coupling constants in Eq. \ref{eq:khg}. 
Note that unlike the fermionic case, the eigenenergies and eigenmodes in this case are obtained by diagonalizing the non-Hermitian matrix $\Sigma \mathcal{M}_{\kk}$, where $\Sigma=$diag$(1,1,-1,-1)$.

In what follows, we shall distinguish two cases motivated by possible directions of external field: (i) spins pointing in $[001]$ direction; (ii) spins pointing in $[111]$ direction. It turns out that the magnon spectrum in the two cases is qualitatively different. 
%For details relating to the spin-wave theory calculations for bothe cases we refer to Ref. \cite{supp}. 
Note that in the absence of the $\gm$ term and no external field, the two cases are degenerate at the classical level. However, they have different zero point energy coming from the quantum correction. The harmonic level correction prefers $[001]$ spin polarization as opposed to spin polarization in the $[111]$ direction. %(see supp. mat. \cite{supp}).

\begin{figure}
\centering
\subfloat[]{\includegraphics[width=0.24\textwidth]{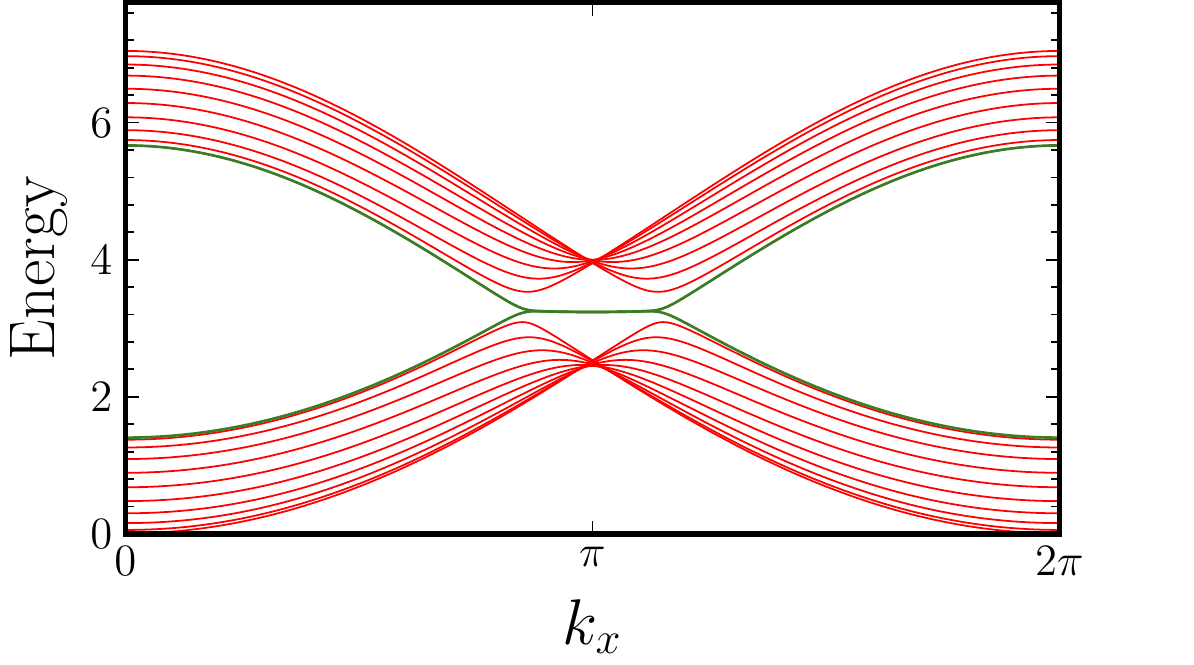}} ~~
\subfloat[]{\includegraphics[width=0.24\textwidth]{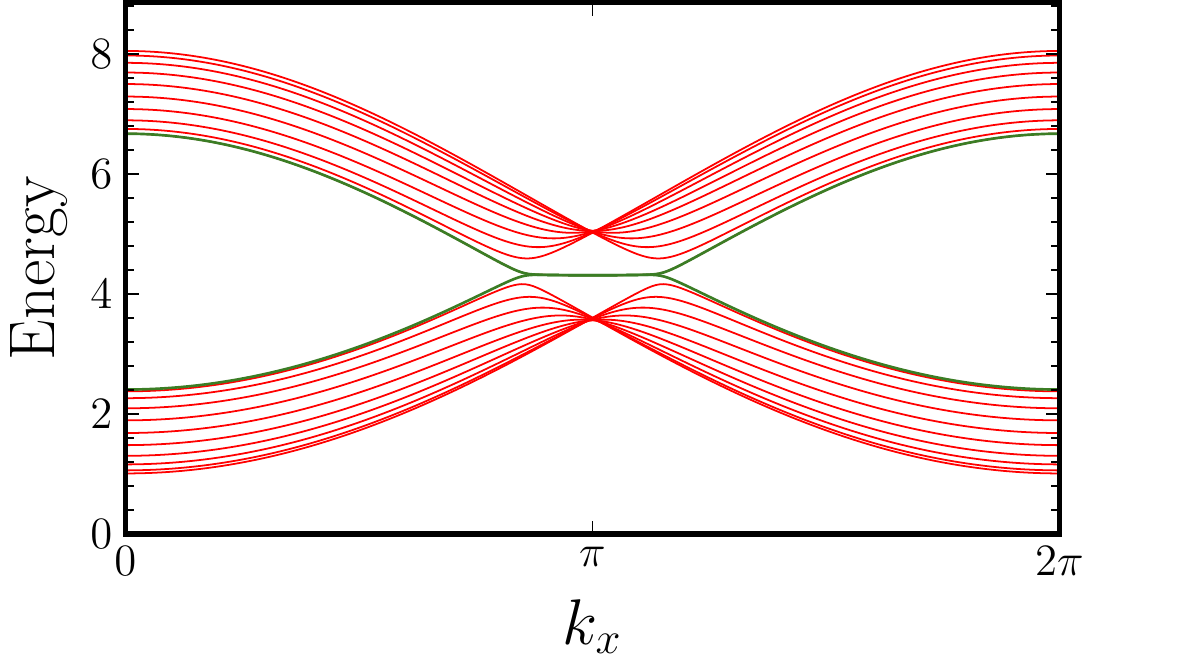}} \\
\subfloat[]{\includegraphics[width=0.24\textwidth]{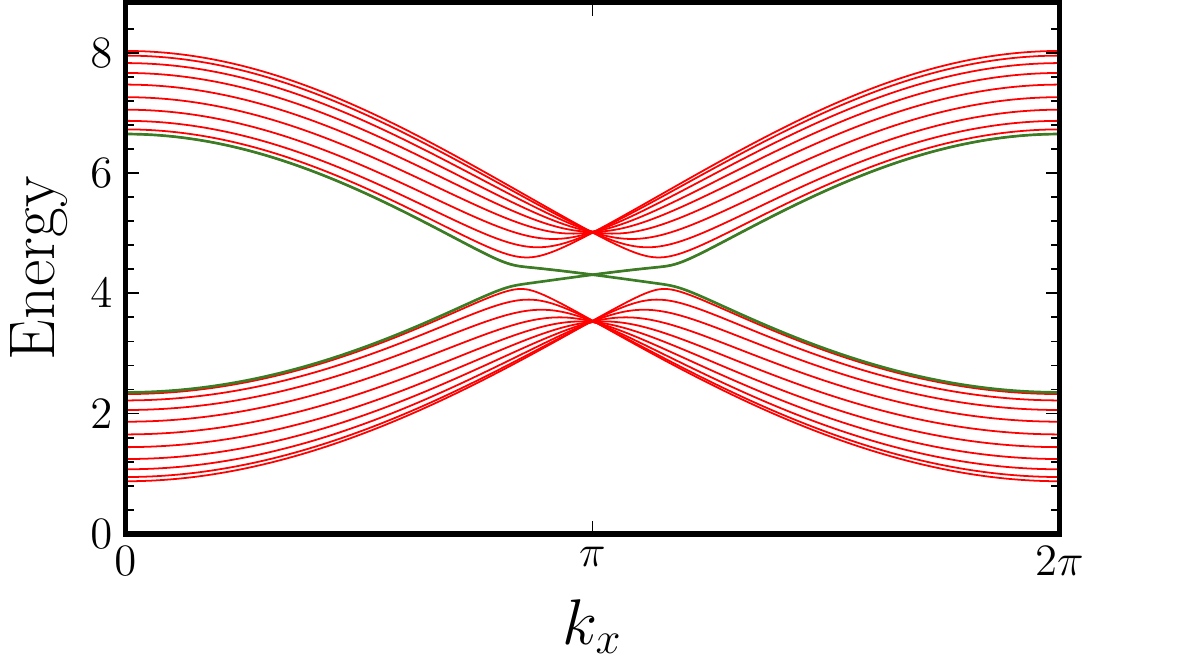}} ~~
\subfloat[]{\includegraphics[width=0.24\textwidth]{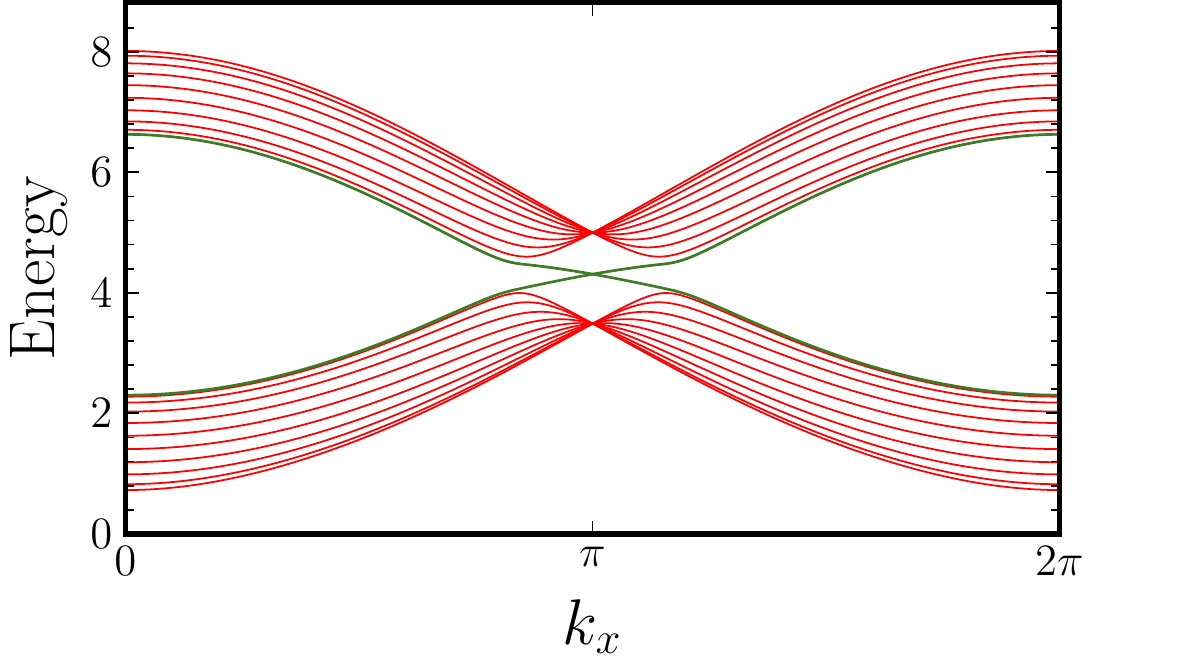}}
\caption{Magnon bands along the zig-zag edge for spin polarization in $[001]$ direction. (a) $h=\gm=0$; (b) $h=0.5$, $\gm=0$; (c) $h=0.5$, $\gm^{z}=0.5$; (d) $h=0.5$, $\gm^{z}=0.7$. In all the plots $\phi=5\pi/4$.}
\label{fig:001_hg}
\end{figure}

\emph{(i) Spin polarization in $[001]$ direction.--}
In this case,
\begin{equation}
\label{eq:AB1}
A_{\kk} =
\left(
\begin{matrix}
\kappa_{0} & \kappa_{1,\kk} \\
\kappa_{1,\kk}^{*} & \kappa_{0}
\end{matrix}
\right) \,; ~~
B_{\kk} =
\left(
\begin{matrix}
0 & \kappa_{2,\kk} \\
\kappa_{2,-\kk} & 0
\end{matrix}
\right) \,,
\end{equation}
where 
\begin{align}
\kappa_{0} &= -3\jh-2\jk+\hz/S \,, \\
\kappa_{1,\kk} &= \jh + (\jh+\jk) \left( e^{-\ii \kk_{1}} + e^{-\ii \kk_{2}} \right) \,, \\
\kappa_{2,\kk} &= \ii \gm^{z} + \jk \left( e^{-\ii \kk_{1}} - e^{-\ii \kk_{2}} \right) \,,
\end{align}
and $\kk_{1,2}=\kk \cdot \vec{a}_{1,2}$ with $\vec{a}_{1,2} = a \lbrace \pm 1/2, \sqrt{3}/2 \rbrace$ ($a$ is the lattice distance between two nearest neighbor A sub-lattice sites, and we shall set $a=1$).
Owing to the two-sublattice structure of the honeycomb lattice, there are two magnon bands. In the absence of both the $\gm$ term and the external magnetic field, these two magnon bands touch each other linearly at the corners of the Brillioun zone. Also, along the zig-zag edge there are non-dispersive edge states connecting these band touching points. The situation is similar to that of graphene, except these are bosons and the band touching is at non-zero energy. Note that the spectrum has Goldstone mode (see fig. \ref{fig:001_hg} (a)).  

\begin{figure}
\centering
\includegraphics[width=0.5\textwidth]{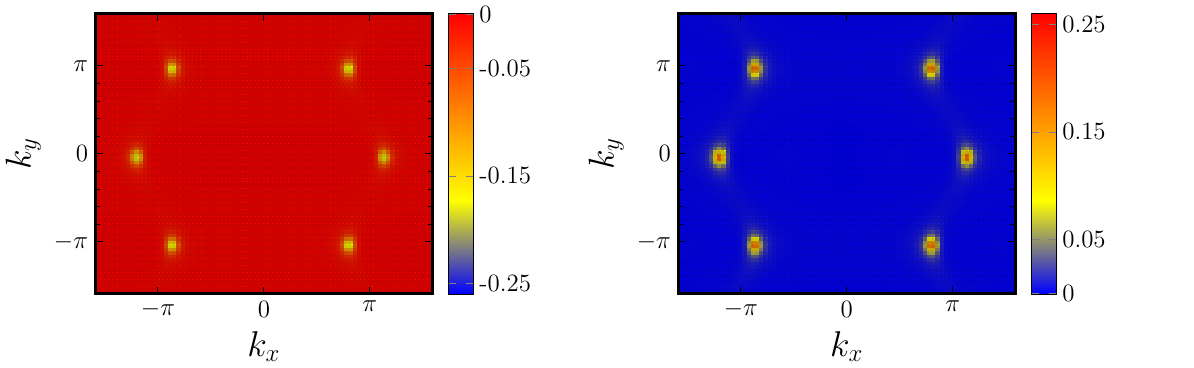}
\caption{Berry curvature for the magnon bands corresponding to fig. \ref{fig:001_hg} (c). This leads to Chern number $\mp 1$ respectively.}
\label{fig:001_berry}
\end{figure}

Introduction of an external field in the $[001]$ direction leads to shifting of the spectrum at higher energy creating a non-zero gap and thus the Goldstone mode is lost. However, a non-zero field in $[001]$ direction does not affect the magnon band touching points (fig. \ref{fig:001_hg} (b)). 
Situation becomes interesting upon introducing a spin-anisotropic interaction along the $z$-bond, i.e. $\gm^{x}=\gm^{y}=0$ and $\gm^{z} \neq 0$. In this case a gap opens between the magnon bands. Remarkably, this gap opening results in dispersive edge states crossing linearly at the band-gap center (see fig. \ref{fig:001_hg} (c) and (d) ). This is analogous to the chiral edge states in the quantum Hall effect. Even in this case, the edge states are protected by a non-zero Chern number topological invariant. 
In fig. \ref{fig:001_berry} we show Berry curvature corresponding to fig. \ref{fig:001_hg} (c). Integrating over the first Brillioun zone we obtain the Chern number $\mp 1$ for the two magnon bands \cite{supp_mat}. 
 
It is important to note that in the absence of an external magnetic field, a non-zero $\gm$ term does not support spin polarization along the $[001]$ direction. %\todo{clarify}. 
We find that upon increasing the field strength in the $[001]$ direction, the gap between the magnon bands reduces and eventually these bands touch each other at high fields. 

%%%%%%%%%%%%%%%%%%%%%%%%%%
%%%%%%%%%%%%%%%%%%%%%%%%%%

%%%%%%%%%%%%%%%%%%%%%%%%%%
%%%%%%%%%%%%%%%%%%%%%%%%%%

\begin{figure}
\centering
\subfloat[]{\includegraphics[width=0.24\textwidth]{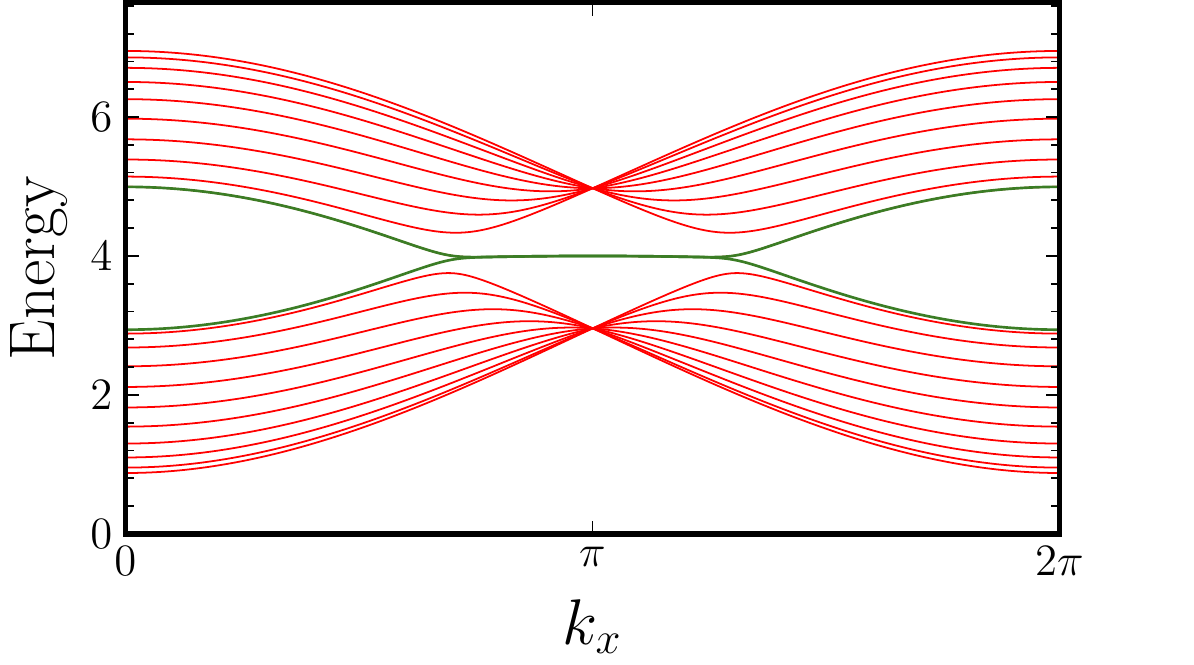}} ~~
\subfloat[]{\includegraphics[width=0.24\textwidth]{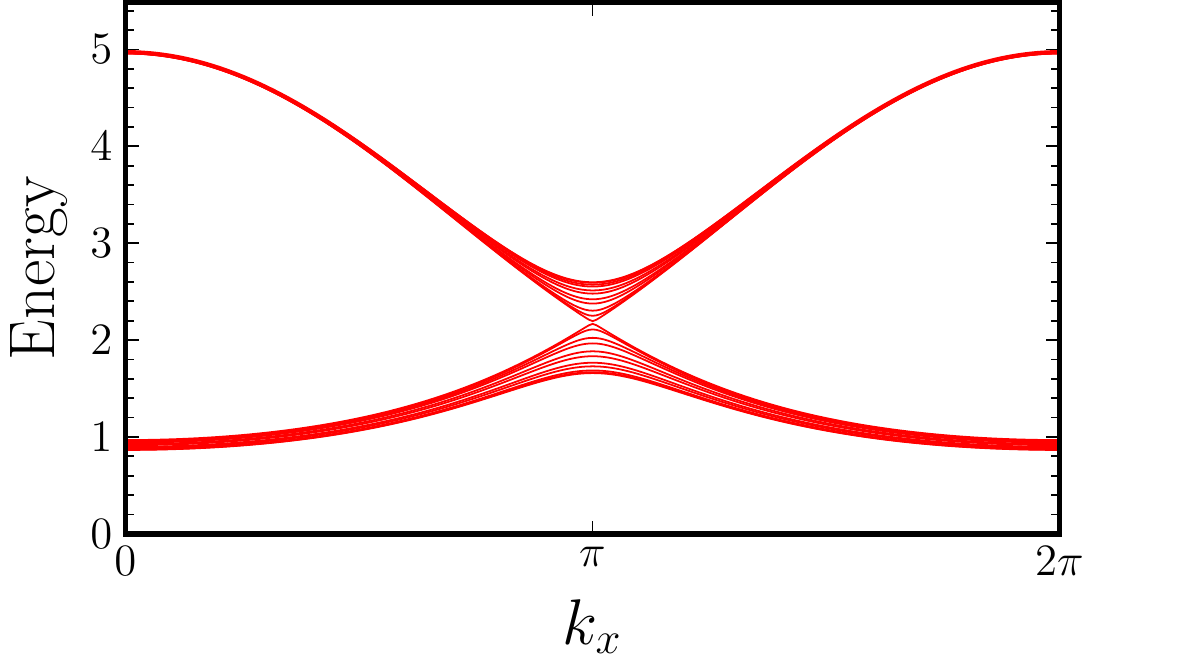}} 
\caption{ Magnon bands along the zig-zag edge for spin polarization in $[001]$ direction. (a) $\phi=\pi$; (b) $\phi=3\pi/2$. In all the plots $h=\gm^{z}=0.5$.
There are no chiral edge states for $\phi=\pi$, which is Heisenberg point. At Kitaev point, the band-touching points merge and so no edge states. }
\label{fig:001_phi}
\end{figure}

On the other hand, increasing the Kitaev interaction towards the Kitaev point leads to flattening of the lower magnon band. Eventually, as $\phi \rightarrow 3\pi/2$ the lower magnon band condenses in extended regions of the Brillioun zone indicating no preferential ordering tendency (see fig. \ref{fig:001_phi}). This is consistent with the expected Kitaev spin liquid across the phase transition. 

Note that a non-zero $\gm^{x}$ and/or $\gm^{y}$ does not support ferromagnetic alignment in $[001]$ direction. %See supplemental material \cite{supp} for details. 
Also, for $\gm^{z}=0$ an external field in $[001]$ direction does not open a gap between the magnon bands.

%%%%%%%%%%%%%%%%%%%%%%%%%%
%%%%%%%%%%%%%%%%%%%%%%%%%%

\begin{figure}
\centering
\subfloat[]{\includegraphics[width=0.24\textwidth]{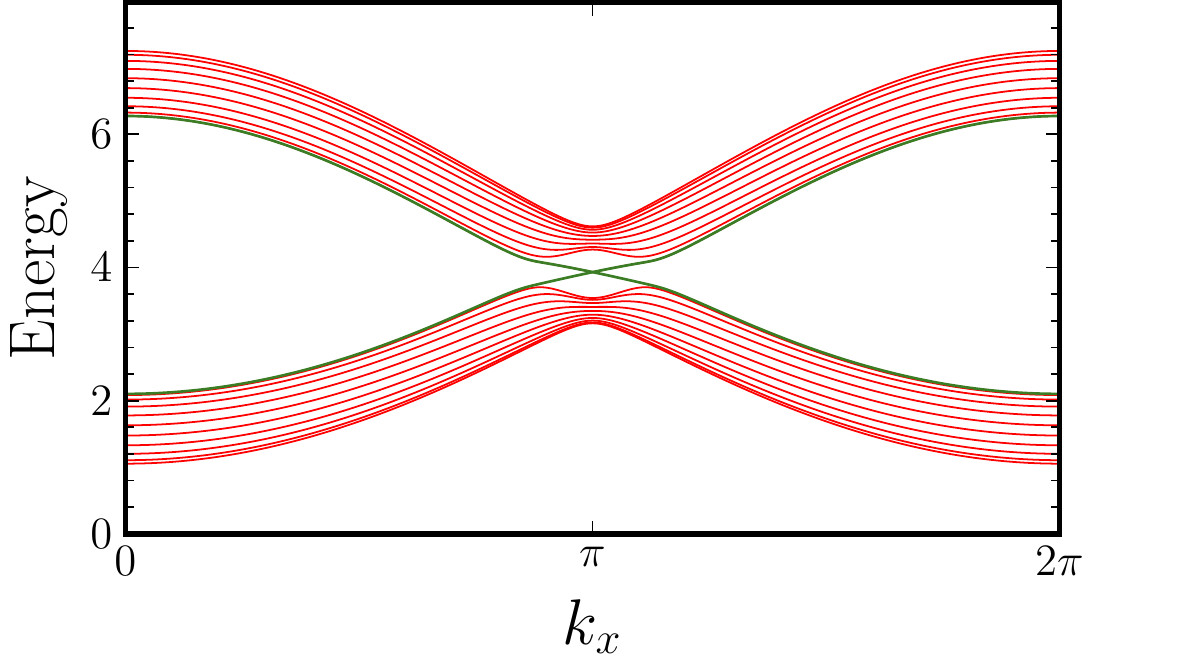}} ~~
\subfloat[]{\includegraphics[width=0.24\textwidth]{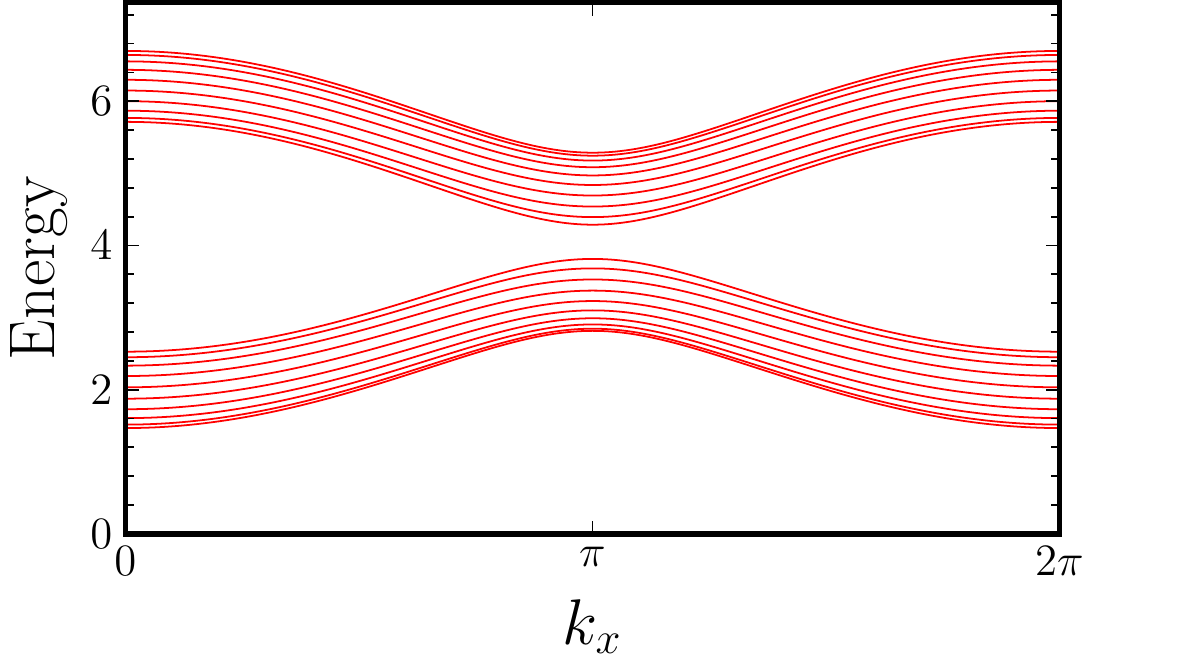}} 
\caption{Magnon bands along the zig-zag edge for spin polarization in $[001]$ direction. (a) $K^{x}=0.8K$; (b) $K^{x}=0.2K$. In all the plots $h=\gm^{z}=0.5$ and $\phi=4\pi/3$.
Tuning of $K^{x}$ leads to magnon band gap closing and reopening, which is accompanied with disappearance of the chiral edge states. }
\label{fig:001_Kx}
\end{figure}

Apart from the external field, there is another route to tune a topological phase transition. Starting from the case with isotropic Kitaev coupling if we make $K^{x,y}$ anisotropic then the band touching points move closer and at a critical value merge and open a band gap. If there are chiral edge states to start with, then upon tuning $K^{x}$ these vanish as the band gap closes and reopens (see fig. \ref{fig:001_Kx}). In real experiments this can be achieved via tuning pressure along the zig-zag edge. 

%%%%%%%%%%%%%%%%%%%%%%%%%%
%%%%%%%%%%%%%%%%%%%%%%%%%%

\begin{figure}
\centering
\subfloat[]{\includegraphics[width=0.24\textwidth]{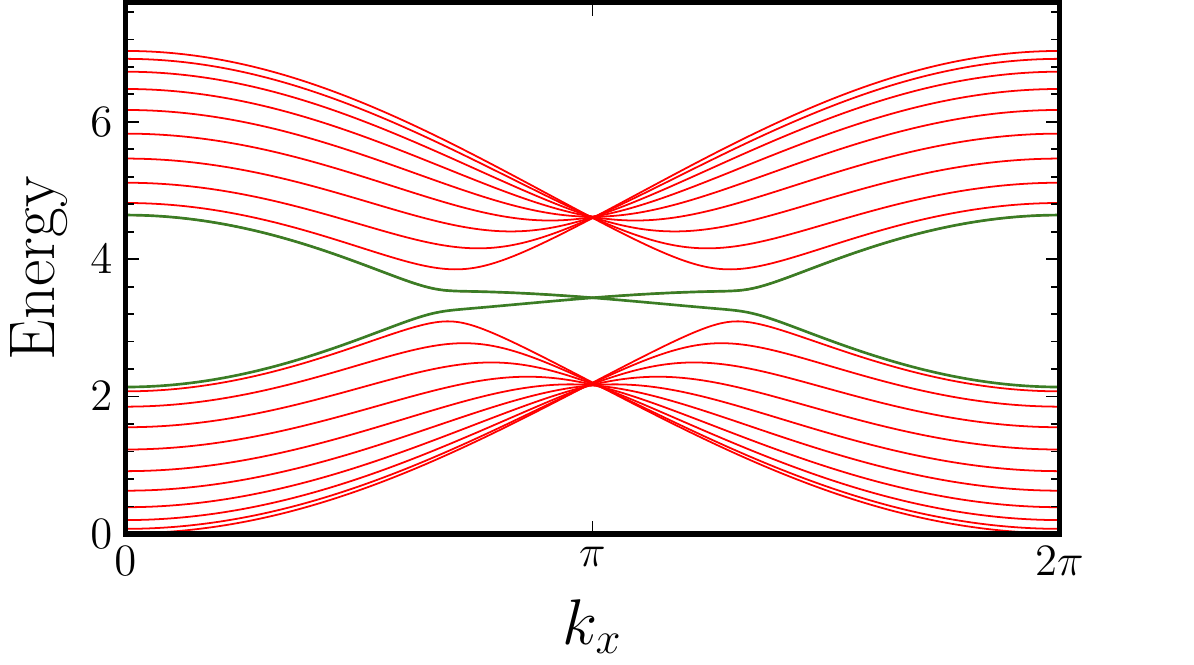}} ~~
\subfloat[]{\includegraphics[width=0.24\textwidth]{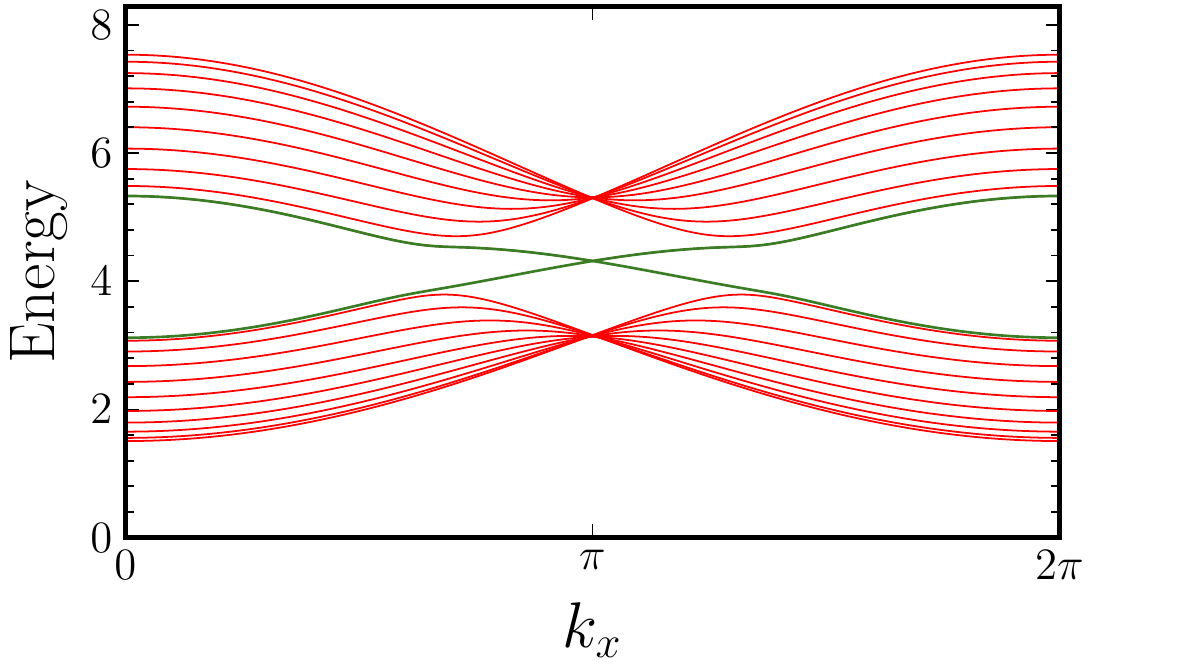}} \\
\subfloat[]{\includegraphics[width=0.24\textwidth]{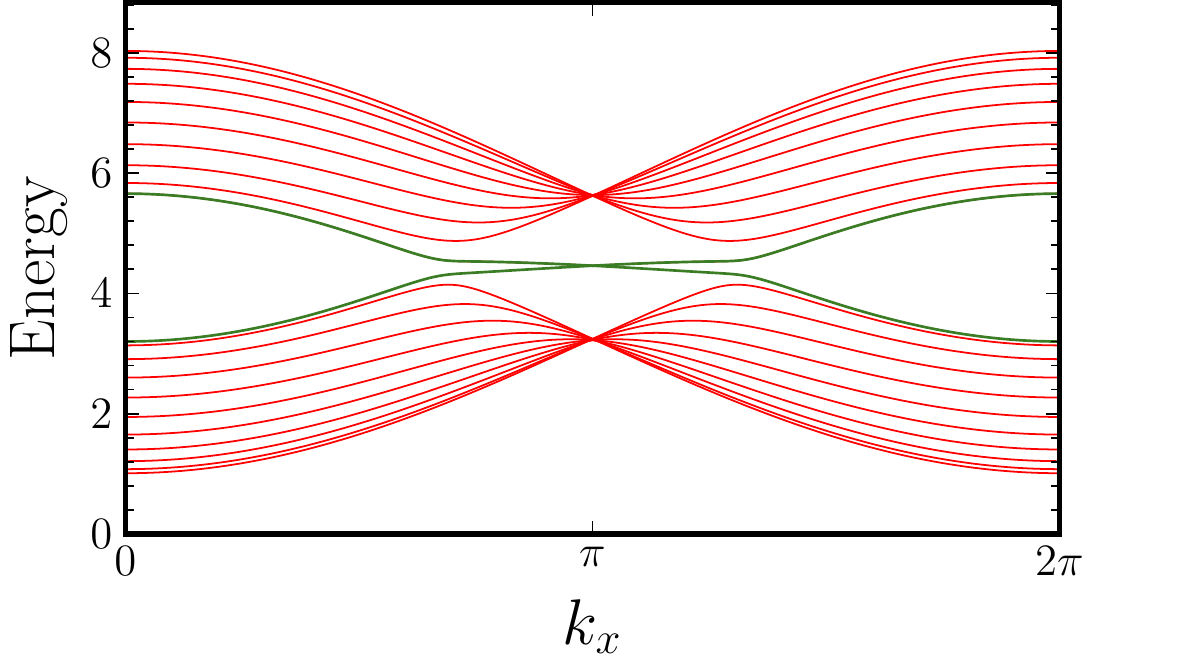}} ~~
\subfloat[]{\includegraphics[width=0.24\textwidth]{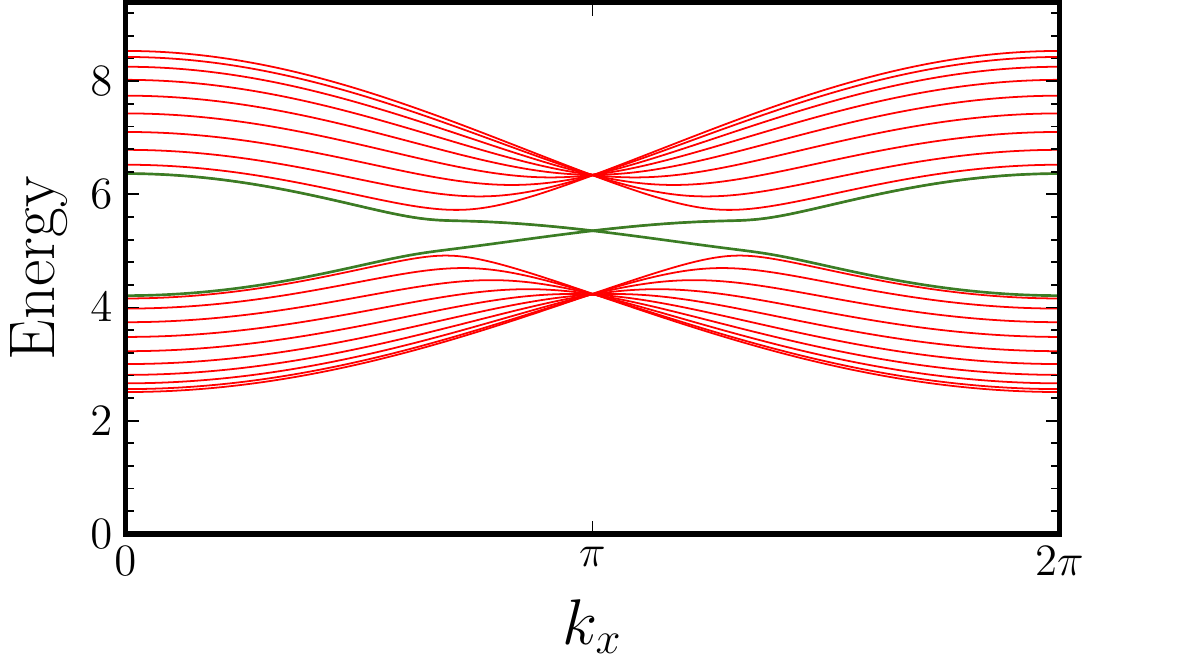}}
\caption{Magnon bands along the zig-zag edge for spin polarization in $[111]$ direction. (a) $h=\gm=0$; (b) $h=0$, $\gm=-0.5$; (c) $h=0.5$, $\gm=0$;
(d) $h=0.5$, $\gm=-0.5$. In all the plots $\phi=5\pi/4$.}
\label{fig:111_hg}
\end{figure}

\emph{(ii) Spin polarization in $[111]$ direction.--}
After suitable rotation in the spin space, we again have
\begin{equation}
\label{eq:AB1}
A_{\kk} =
\left(
\begin{matrix}
\kappa_{0} & \kappa_{1,\kk} \\
\kappa_{1,\kk}^{*} & \kappa_{0}
\end{matrix}
\right) \,; ~~
B_{\kk} =
\left(
\begin{matrix}
0 & \kappa_{2,\kk} \\
\kappa_{2,-\kk} & 0
\end{matrix}
\right) \,,
\end{equation}
where 
\begin{align}
\kappa_{0} &= -3\jh-2\jk-2\gm+\hz/S \,, \\
\kappa_{1,\kk} &= \left(\jh+ \frac{2\jk}{3} - \frac{\gm}{3} \right) \left(1+ e^{-\ii \kk_{1}} + e^{-\ii \kk_{2}} \right) \,, \\
\kappa_{2,\kk} &= \left(\frac{2\jk}{3} + \frac{2\gm}{3} \right) \left(1+ e^{-\ii (\kk_{1}+2\pi/3)} + e^{-\ii (\kk_{2}-2\pi/3)} \right) \,.
\end{align}
%For the spin polarization along the $[111]$ in the absence of field and $\gm$ term, the situation is similar to that discussed %above. However, in this case, to open a gap between the magnon bands we need $\gm^{x}=\gm^{y}=\gm^{z}\neq 0$ in the absence of %an external field. With the introduction of these gap opening terms we immediately obtain chiral edges states along the zig-zag %edge of the honeycomb lattice. 
For the spin polarization along the $[111]$ direction in the absence of field and $\gm$ term, there is Goldstone mode as well as magnon band gap along with chiral edge states (see Fig. \ref{fig:111_hg} (a)). However, as stated before, the $[111]$ spin polarization is not favored by the harmonic level zero-point energy for $\hz=\gm=0$. Thus we do not discuss this case. Note that anisotropic $\gm$ terms do not support $[111]$ spin polarization. For any $\gm^{x}=\gm^{y}=\gm^{z}\neq 0$ in the absence of external field, there is magnon band gap along with chiral edge states along the zig-zag edge of the honeycomb lattice 
(see Fig. \ref{fig:111_hg} (b)).  

Upon adding an external magnetic field in the $[111]$ direction the gap between the magnon bands again closes at high fields. However, what is interesting to note here is that even in the absence of the $\gm$ term, the external field alone is sufficient to open a gap between the magnon bands and produce chiral edge states along the zig-zag edge (see fig. \ref{fig:111_hg} (c)). 

\begin{figure}
\centering
\includegraphics[width=0.5\textwidth]{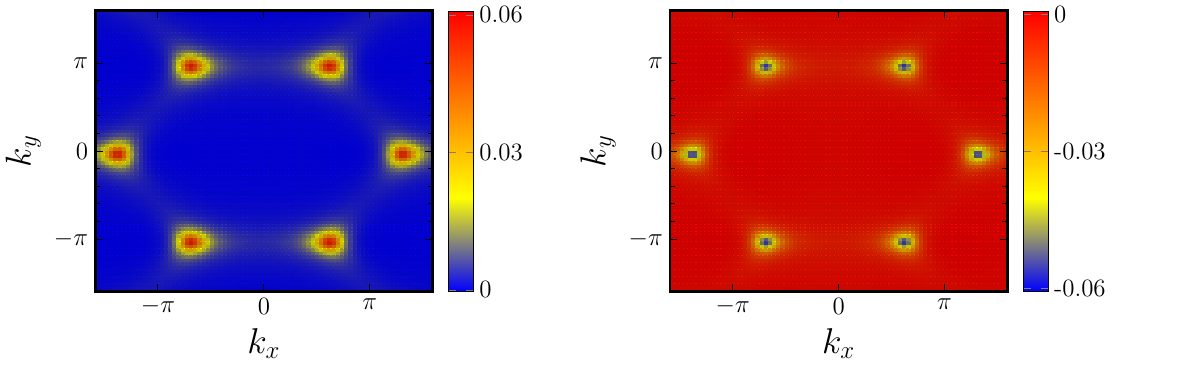}
\caption{Berry curvature for the magnon bands corresponding to Fig. \ref{fig:111_hg} (b). This leads to Chern number $\pm 1$ respectively.}
\label{fig:111_berry}
\end{figure}

The chiral edge states discussed here are again topologically protected via a non-zero Chern number. In fig. \ref{fig:111_berry}, we show Berry curvature corresponding to Fig. \ref{fig:111_hg} (b), which gives Chern number $\pm 1$ for the two magnon bands. 

Upon increasing the Kitaev term so that we approach the ferromagnetic Kitaev point, the lower magnon band flattens and approaches zero energy. At the phase transition we have condensation of magnons in extended regions in the Brillioun zone suggesting no particular ordering tendency as a pre-cursor to the Kitaev spin liquid. 

Note that in this case it is not clear whether the ferromagnetic phase with $[111]$ spin polarization is stable in case of anisotropic Kitaev couplings. Hence this route to tune a topological phase transition, unlike in case (i), is absent here. 

Another interesting feature is that even at $\phi=\pi$, i.e. the Heisenberg point, a small $\gm$ term in presence of field opens the magnon band gap to give chiral edge states. Without the $\gm$ term, the magnon bands touch each other even in presence of finite field. This is different from the case (i), wherein even finite $\gm$ and finite field do not open the magnon band gap at the Heisenberg point. In case (i), the Kitaev interaction is essential to realize chiral edge states.

%%%%%%%%%%%%%%%%%%%%%%%%%%
%%%%%%%%%%%%%%%%%%%%%%%%%%

%%%%%%%%%%%%%%%%%%%%%%%%%%
%%%%%%%%%%%%%%%%%%%%%%%%%%

\emph{Conclusions.--}
In this work we have shown that there are topologically non-trivial excitations in the ferromagnetic phase of the extended Kitaev-Heisenberg model. We have discussed two cases depending on the direction of the spin polarization and shown that both the cases host bosonic topological excitations at the edges. It is clear that the spin-anisotropic interaction is responsible to open a gap between the magnon bands. While in the absence of $\gm$ term a gap opening is possible by applying an external field in $[111]$ direction, a field in $[001]$ direction alone is not sufficient in gap opening. Most importantly, there are chiral edge states upon opening the magnon band gap in both the cases. These edge states are topologically protected by a non-zero Chern number. 

Although our analysis is based on linear spin-wave theory, magnon interactions are not expected to alter the discussed Physics given the fact that there is a systematic expansion parameter $(1/S)$. Interactions will renormalize the magnon dispersion, mostly likely to flatten the magnon bands as in most frustrated systems. However, the magnon band gap is not expected to close. 

\begin{figure}[ht]
\centering
\subfloat[]{\includegraphics[width=0.24\textwidth]{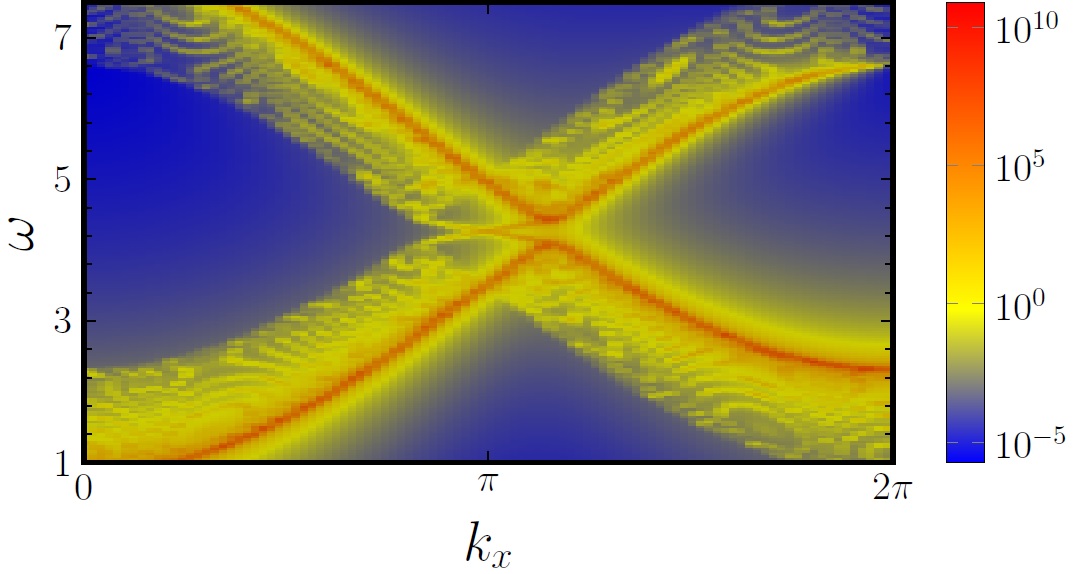}} ~~
\subfloat[]{\includegraphics[width=0.24\textwidth]{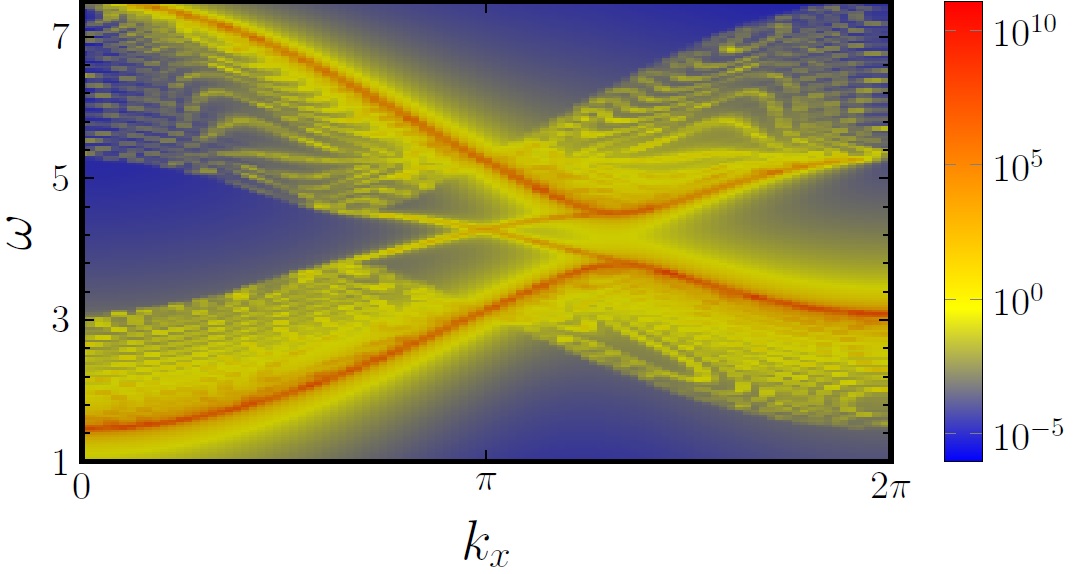}} 
\caption{Dynamic structure factor at $k_y =0$ corresponding to (a) Fig. \ref{fig:001_hg} (c) and (b) Fig. \ref{fig:111_hg} (b). The edge states along the zig-zag edge can be seen around $k_x=\pi$ along with the bulk magnon bands. 
%(a) Spin polarization in $[001]$ direction with $h=0.5$, $\gm^{z}=0.5$; 
%(b) Spin polarization in $[111]$ direction with $h=0$, $\gm=-0.5$. In all the plots $\phi=5\pi/4$.
}
\label{fig:susc}
\end{figure}

In principle, neutron scattering experiments can detect these topological edge excitations. The dynamic structure factor shows signatures of these edge states \cite{supp_mat}, as shown in fig. \ref{fig:susc}. However,
it is important to keep in mind that interaction induced decay is possible \cite{decay_review} and that the edge states are susceptible to it \cite{topo_decay}. In particular, due to the presence of cubic terms in the Hamiltonian in case (ii) a two-particle decay is possible when the spin polarization is in the $[111]$ direction. This means that spectroscopic experiments might not be the best probe to study these topological excitations. Although, we must point out that at higher fields due to phase-space constraints, the interaction-induced decay may be restricted.

%%%%%%%%%%%%%%%%%%%%%%%%%%
%%%%%%%%%%%%%%%%%%%%%%%%%%

A promising technique to detect topological excitations in quantum magnets seems to be spin-Hall noise spectroscopy, which is based on inverse spin-Hall effect. Since it depends directly on the edge-spin correlations only, the signal from the edge states is significantly enhanced as opposed to the bulk magnons \cite{shns}. Moreover, thermal transport \cite{heat} and spin current \cite{spin_trans} are viable detection techniques.

Our work opens a whole new set of interesting theoretical questions. In particular, it is important to investigate whether other ordered phases in the Kitaev-Heisenberg model also host topological excitations. We plan to address this question in future work. This also means that it is worth putting experimental efforts in detecting topological excitations in candidate Kitaev materials.  

\emph{Note.--} 
During the final stages of this work Ref. \cite{kh_topo} appeared, which studies topological excitations at high magnetic fields in $[111]$ direction, both analytically and numerically. Our findings of case (ii) in presence of a $[111]$ field are consistent with those in Ref. \cite{kh_topo}. 

% \emph{Acknowledgments.---}
I acknowledge discussion with Andreas Schnyder and Matthias Vojta. I thank Andreas Schnyder for reading the manuscript.

\bibliographystyle{apsrev}
\bibliography{bib_kh_topo_arxiv_v3}
 
%%%%%%%%%%%%%%%%%%%%%%%%%%%%%%%%%%%%%%%%%%%%%%%%%%
%%%%%%%%%%%%%%%%%%%%%%%%%%%%%%%%%%%%%%%%%%%%%%%%%%
 
%%%%%%%%%%%%%%%%%%%%%%%%%%%%%%%%%%%%%%%%%%%%%%%%%%
%%% Supp. Mat.
%%%%%%%%%%%%%%%%%%%%%%%%%%%%%%%%%%%%%%%%%%%%%%%%%%

%%%%%%%%%%%%%%%%%%%%%%% Supplemental material %%%%%%%%%%%%%%%%%%%%%%%%%%%%%
\clearpage 
%\newpage

%%%%%%%%%%
\setcounter{equation}{0}
\setcounter{figure}{0}
\setcounter{table}{0}
\setcounter{page}{1}
\renewcommand{\theequation}{S\arabic{equation}}
\renewcommand{\thefigure}{S\arabic{figure}}
\renewcommand{\bibnumfmt}[1]{[S#1]}
\renewcommand{\citenumfont}[1]{S#1}
%%%%%%%%%%
%\onecolumn
\onecolumngrid
\begin{center}
\textbf{\large Supplemental material: \\
Topological excitations in the ferromagnetic Kitaev-Heisenberg model} \\
\medskip 

Darshan G. Joshi  \\
\smallskip

{\em Max-Planck-Institute for Solid State Research, D-70569 Stuttgart, Germany}

\end{center}
%\end{widetext}

\bigskip

\twocolumngrid
%%%%%%%%%%%%%%%%%%%%%%%%%%%%%%%%%%%
%%%%%%%%%%%%%%%%%%%%%%%%%%%%%%%%%%%

%%%%%%%%%%%%%%%%%%%%%%%%%%%%%%%%%%%
\section{Dynamic structure factor}

Neutron scattering experiments can access the dynamic structure factor. Here we shall calculate it for the extended Kitaev-Heisenberg model (Eq. (1) in main text) on the honeycomb lattice with open boundaries to see the signature of the Chiral edge states along the zig-zag edge. Since the crystal momentum in $y-$direction is not a good quantum number anymore, we calculate the $k_y=0$ contribution in a scattering experiment. The dynamic structure factor,
\begin{equation}
\label{eq:sf_def}
S(\kk,\omega) = \frac{1}{N} \sum_{i,j} S_{ij} e^{\kk \cdot \vec{r}_{ij}} \,,
\end{equation}
%So,
%\begin{equation}
%\label{eq:s_w}
%S(k_x,\omega) \equiv S(k_y=0,\omega) = \frac{1}{N} \sum_{i,j} S_{ij} \,,
%\end{equation}
where $S_{ij} = - \Im \chi_{ij}$, with $\chi$ being the spin correlation function and $\Im$ stands for the imaginary part. We shall focus on the $k_y=0$ contribution with $k_x$ momentum dependence still present. Recall that we have to perform a bosonic Bogoliubov transformation, i.e. diagonalize a non-Hermitian matrix, in order to obtain the eigen modes. As a result, the original magnon operators $b$ are represented in terms of the Bogoliubov quasiparticle operators $\bt$ as follows:
\begin{align}
\label{eq:bg}
b_{A,i} &= \sum_{m=1}^{N} \bigg[ u_{i,m} \bt_{A,m} + u_{i,m+N} \bt_{B,m}  \nonumber \\ 
&~~~~~~~~~+ v_{i,m} \bt^{\dagger}_{A,m} + v_{i,m+N} \bt^{\dagger}_{B,m} \bigg] \,, \\
b_{B,i} &= \sum_{m=1}^{N} \bigg[ u_{i+N,m} \bt_{A,m} + u_{i+N,m+N} \bt_{B,m} \nonumber \\ 
&~~~~~~~~~+ v_{i+N,m} \bt^{\dagger}_{A,m} + v_{i+N,m+N} \bt^{\dagger}_{B,m} \bigg] \,.
\end{align}
Here the index $i$ on the LHS is used to denote each different zig-zag stripe along the $y-$direction with $k_x$ momentum dependence implicitly present. We consider $N$ such stripes with open boundary condition and hence there are $2N$ modes.  
Also, $u_{i,j}$ and $v_{i,j}$ are matrix elements of the $2N \times 2N$ Bologliubov transformation matrices $U$ and $V$ respectively. In terms of these matrices, the Hamiltonian is diagonalized in the following way:
\begin{equation}
\label{eq:diag}
\Omega = T^{\dagger} \mathcal{H} T \,,
\end{equation}
where $\Omega$ is the diagonal matrix with eigenmodes of the Hamiltonian matrix $\mathcal{H}$. The $4N \times 4N$ matrix,
\begin{equation}
\label{eq:t}
T = \left(
\begin{matrix}
U & V \\
V^{*} & U^{*} 
\end{matrix}
\right) \,,
\end{equation}
such that it satisfies the condition, $T^{\dagger} \Sigma T = T \Sigma T^{\dagger} = \Sigma$. $\Sigma =$diag$(\mathbbm{1},\mathbbm{1},-\mathbbm{1},-\mathbbm{1})$ where $\mathbbm{1}$ is $2N \times 2N$ identity matrix.

\begin{figure}
\centering 
\subfloat[]{\includegraphics[width=0.25\textwidth]{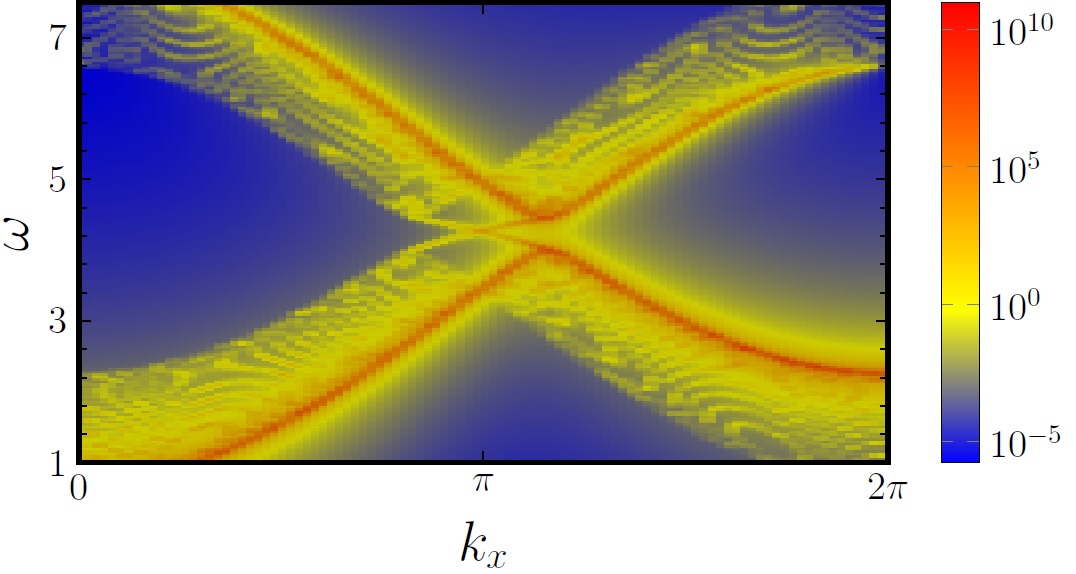}} ~
\subfloat[]{\includegraphics[width=0.25\textwidth]{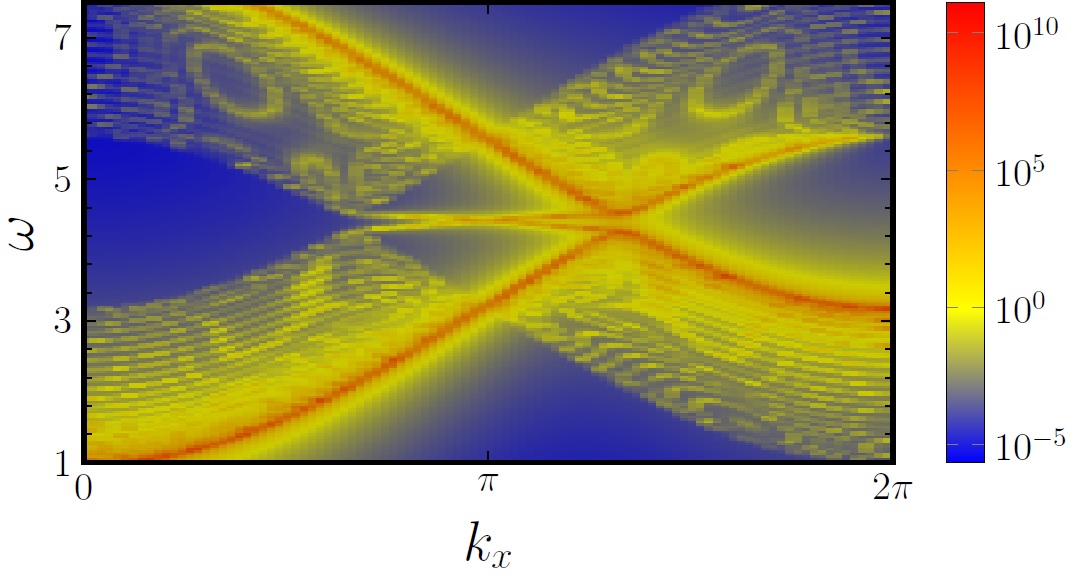}} \\
\subfloat[]{\includegraphics[width=0.25\textwidth]{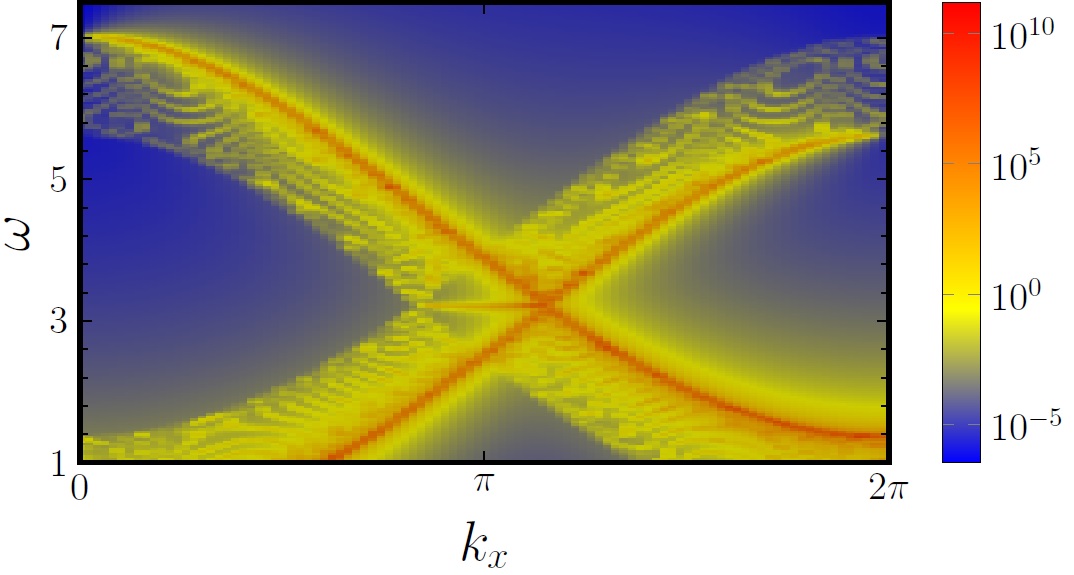}} ~
\subfloat[]{\includegraphics[width=0.25\textwidth]{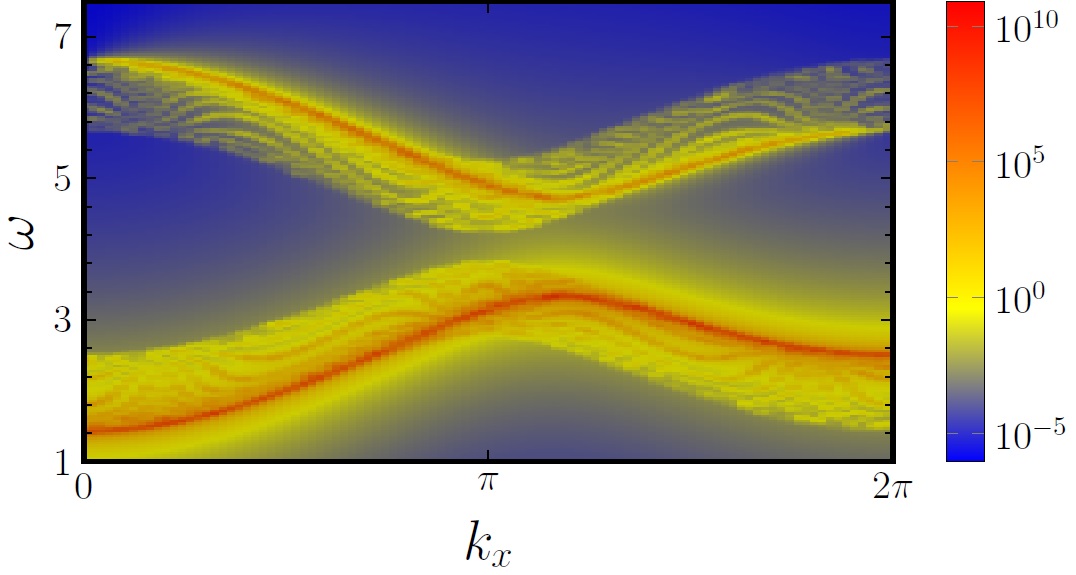}}
\caption{Dynamic structure factor for different parameters and spin polarization, with zig-zag edge. 
(a) Chiral edge states; spin polarization in $[001]$ direction, with $\hz=0.5$, $\gm^{z}=0.7$, and $\phi=5\pi/4$. 
(b) Chiral edge states; spin polarization in $[111]$ direction, with $\hz=0.5$, $\gm=0$, and $\phi=5\pi/4$.
(c) Flat edge states; spin polarization in $[001]$ direction, with $\hz=0$, $\gm=0$, and $\phi=5\pi/4$. 
(d) No edge states; spin polarization in $[001]$ direction, with $\hz=0.5$, $\gm^{z}=0.5$, $\jk^{x}=0.2\jk$ and $\phi=4\pi/3$.}
\label{fig:dsf}
\end{figure}

After performing the Bogoliubov transformation and a few steps of algebra, we obtain,
\begin{align}
\label{eq:ss}
S_{ij} &= \sum_{m=1}^{N} \bigg \lbrace  
\delta(\omega - \omega_{A,m})  \nonumber \\
&\times \bigg[  
u_{i,m} u^{*}_{j,m} + v^{*}_{i,m} v_{j,m} + u_{i+N,m} u^{*}_{j+N,m} + v^{*}_{i+N,m} v_{j+N,m} \nonumber \\
&+ u_{i,m} u^{*}_{j+N,m} + v^{*}_{i,m} v_{j+N,m} + u_{i+N,m} u^{*}_{j,m} + v^{*}_{i+N,m} v_{j,m} 
\bigg] \nonumber \\
&+ \delta(\omega - \omega_{B,m}) \nonumber \\
&\times \bigg[  
u_{i,m+N} u^{*}_{j,m+N} + v^{*}_{i,m+N} v_{j,m+N} + u_{i+N,m+N} u^{*}_{j+N,m+N}  \nonumber \\
&+ v^{*}_{i+N,m+N} v_{j+N,m+N} + u_{i,m+N} u^{*}_{j+N,m+N} + v^{*}_{i,m+N} v_{j+N,m+N} \nonumber \\
&+ u_{i+N,m+N} u^{*}_{j,m+N} + v^{*}_{i+N,m+N} v_{j,m+N}  
\bigg]
\bigg \rbrace \,.
\end{align} 
Inserting this into Eq. \ref{eq:sf_def} we obtain the dynamic structure factor. This is shown in Fig. \ref{fig:dsf} as well as  in Fig. 8 in the main text. Note that we have added a Lorentzian broadening $\lambda = 10^{-3}$ to the above delta functions while plotting. 

%%%%%%%%%%%%%%%%%%%%%%%%%%%
%%%%%%%%%%%%%%%%%%%%%%%%%%%

As seen in Fig. \ref{fig:dsf} (a)--(b), the dynamic structure factor clearly shows spectral weight around $k_x=\pi$ for the chiral edge states with linear dispersion. In contrast, in Fig. \ref{fig:dsf} (c) we see that the chiral edge states are absent and only flat non-dispersing edge states are present. These signatures are absent in Fig. \ref{fig:dsf} (d), where there are no edge states. In real neutron scattering experiments, however, there will be several effects such as interaction-induced decay as well as thermal broadening which might diminish the signal of edge states compared to the bulk. However, spin-Hall noise spectroscopy, which depends only on the edge-spin correlation is likely to have significant contribution from the edges compared to the bulk.

%%%%%%%%%%%%%%%%%%%%%%%%%%%%%%%%%%%

%%%%%%%%%%%%%%%%%%%%%%%%%%%%%%%%%%%

\section{Berry curvature and Chern number}

In order to establish the topological nature of the edge excitations, we shall calculate the Berry curvature and consequently show that it leads to a non-zero Chern number in case of chiral edge states. While there is a straightforward formula to calculate the Chern number for tight-binding Dirac Hamiltonians, presence of anomalous terms in our case complicates the situation. 

The Chern number for a Bloch band with normalized wavefunction $|n(\kk)\rangle$ is given by \cite{chern_s, berry_s, simon_s},
\begin{equation}
\label{eq:chern}
\mathcal{C}_{n} = \frac{1}{2\pi \iota} \int_{BZ} d \kk F_{12}(\kk) \,,
\end{equation}
where $F_{12}(\kk)$ is the Berry curvature defined in terms of the Berry connection, $A_{\mu}(\kk) = \langle n(\kk) |\partial_{\mu} | n(\kk) \rangle$, as follows:
\begin{equation}
\label{eq:bcur}
F_{12}(\kk) = \partial_{1} A_{2}(\kk) - \partial_{2} A_{1}(\kk) \,,
\end{equation}
where $\partial_{\mu}$ is the partial derivative with respect to the $\mu-$component $(\mu=1,2)$ of the momentum vector. The above formula is useful in continuum but not very efficient to implement on a discrete lattice. We therefore use the discrete version to calculate the Berry curvature and the Chern number, as detailed in Ref. \cite{fukui_s}. 

\begin{figure}
\centering
\subfloat[]{\includegraphics[width=0.5\textwidth]{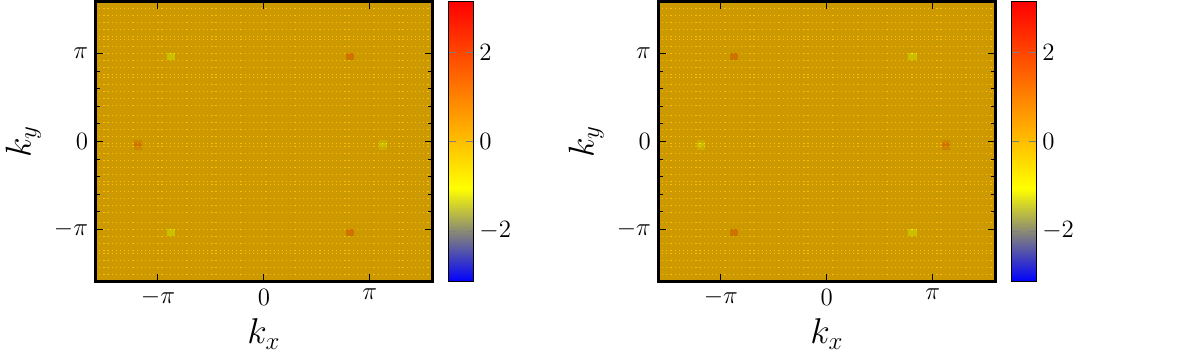}} \\
\subfloat[]{\includegraphics[width=0.5\textwidth]{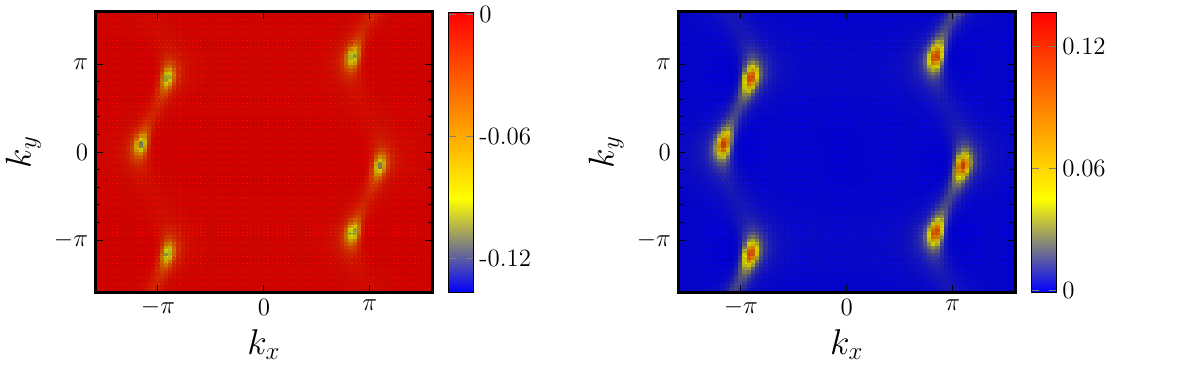}} \\
\subfloat[]{\includegraphics[width=0.5\textwidth]{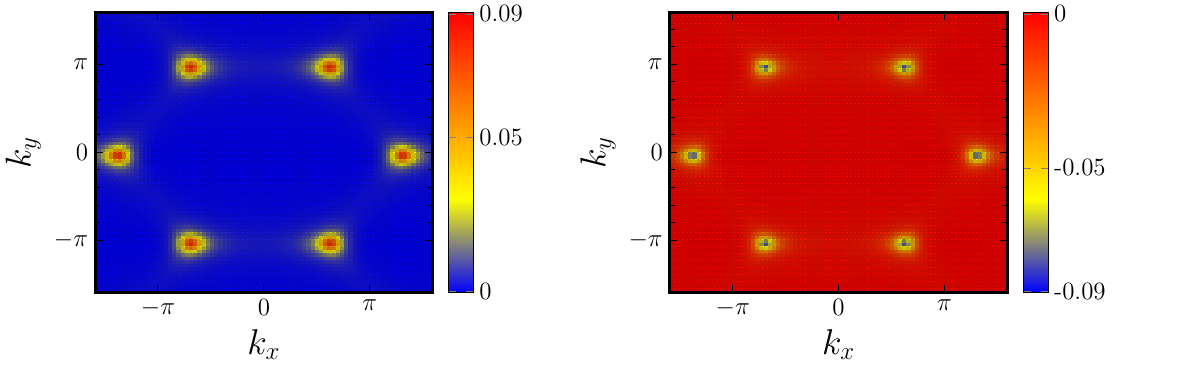}} 
\caption{Berry curvature for the two magnon bands. 
(a) Spin polarization in $[001]$ direction; $\hz=0.5$, $\gm=0$, and $\phi=5\pi/4$. 
(b) Spin polarization in $[001]$ direction; $\hz=\gm^{z}=0.5$, $\jk^{x}=0.8$, and $\phi=4\pi/3$. 
(c) Spin polarization in $[111]$ direction; $\hz=0.5$, $\gm=-0.5$, and $\phi=5\pi/4$. }
\label{fig:berry_plus}
\end{figure}

The lattice Berry curvature is defined as follows \cite{fukui_s}:
\begin{equation}
\label{eq:bcur_lat}
F_{12}(k_{\mu}) \equiv \ln W_{1} (k_{\mu}) W_{2} (k_{\mu} + \hat{1}) W_{1}^{-1} (k_{\mu}+\hat{2}) W_{2}^{-1} (k_{\mu})
\end{equation}
such that $-\iota\pi < F_{12}(k_{\mu}) < \iota\pi$, with $\hat{1} \equiv \hat{k_{x}}$ and $\hat{2} \equiv \hat{k_{y}}$. The link variable $W$ is defined as follows:
\begin{equation}
\label{eq:w_def}
W_{\nu}(k_{\mu}) \equiv \frac{\Phi_{\nu}(k_{\mu})}{|\Phi_{\nu}(k_{\mu})|} \,,
\end{equation}
where $\Phi_{\nu}(k_{\mu})=\langle \phi (k_{\mu}) | \Sigma | \phi (k_{\mu} + \hat{\nu}) \rangle$, and $\phi = \left( U V^{*} \right)^T$ is the eigenmode corresponding to eigenenergy $\omega_n$. Note that since the bosonic eigenmodes obtained after the Bogoliubov transformation are normalized with respect to the $\Sigma$ matrix we have the modified expression for $\Phi$, else for normalized wavefunctions it simply involves usual inner product. Once we calculate the Berry curvature it is then straightforward to obtain the lattice Chern number,
\begin{equation}
\mathcal{C} = \frac{1}{2 \pi \iota} \sum_{\mu} F_{12} (k_{\mu}) \,.
\end{equation}
We have used the above expressions to plot the Berry curvature in Fig. 3 and Fig. 7 in the main text. In Fig. \ref{fig:berry_plus} we plot Berry curvature for some more parameters. For instance, in Fig. \ref{fig:berry_plus} (a) we see that the neighboring pairs of magnon band touching points have opposite Berry curvature. Thus the Chern number is zero, but there is flat edge state connecting these points with opposite Berry curvature (see also Fig. 2 (b) in the main text). Whereas, in Fig. \ref{fig:berry_plus} (b) we see that the Berry curvature of both the bands is non-zero and that the magnon band touching points are moving close to each other. This non-zero Berry curvature then leads to a non-zero Chern number and consequently chiral edge states (see also Fig. 5 (a) in the main text). Similar scenario is seen in Fig. \ref{fig:berry_plus} (c).

%%%%%%%%%%%%%%%%%%%%%%%%%%%%%%%%
%%%%%%%%%%%%%%%%%%%%%%%%%%%%%%%%

Integrating the Berry curvature gives us the Chern number. Note that in order to implement the procedure in Ref. \cite{fukui_s} it is required to use a square Brillioun zone such that the discretization involves mesh with square building blocks. This is easily achieved by considering a rectangular Brillioun zone and then appropriately scaling one of the sides. In our case we considered a grid of $100 \times 100$ points for $-2\pi \leq k_{x} \leq 2\pi$ and $-4\pi/\sqrt{3} \leq k_{y} \leq 4\pi/\sqrt{3}$. We then scale the $k_y$ momentum by $2\pi/\sqrt{3}$ to obtain a square mesh. We have checked our results for different grid sizes and it is unaffected within numerical accuracy.

%%%%%%%%%%%%%%%%%%%%%%%%%%%%%%%%%%%%

%\bibliographystyle{apsrev}
%\bibliography{bib_kh_topo_arxiv_v3}

%%%%%%%%%%%%%%%%%%%%%%%%%%%%%%%%%%%%%%%%%%%%%%%%%%%%%%%%%%%%%%%%%%%%%%%%%%%%%%%

\end{document}